\documentclass[review]{elsarticle}

\usepackage[bookmarks=false]{hyperref}
\usepackage{textgreek}
\usepackage{graphicx} 
\usepackage{xcolor,soul}

\journal{Journal of Theoretical Biology}

\begin{document}

\begin{frontmatter}

\title{A mathematical modelling portrait of Wnt signalling in early vertebrate embryogenesis}

\author{Claudiu V. Giuraniuc\corref{mycorrespondingauthor}}
\cortext[mycorrespondingauthor]{Corresponding author}
\ead{v.c.giuraniuc@abdn.ac.uk}
\author{Shabana Zain}
\author{Shahmama Ghafoor}
\author{Stefan Hoppler}
\address{Institute of Medical Sciences, Foresterhill Health Campus, University of Aberdeen, Aberdeen AB25 2ZD Scotland, UK}

\begin{abstract}
There are two phases of Wnt signalling  in early vertebrate embryogenesis: very early, maternal Wnt signalling promotes dorsal development, and slightly later, zygotic Wnt signalling promotes ventral and lateral mesoderm induction. However, recent molecular biology analysis has revealed more complexity among the direct Wnt target genes, with at least five classes. Here in order to test the logic and the dynamics of a new Gene Regulatory Network model suggested by these discoveries we use mathematical modelling based on ordinary differential equations (ODEs). Our mathematical modelling of this Gene Regulatory Network reveals that a simplified model, with one "super-gene" for each class is sufficient to a large extent to describe the regulatory behaviour previously observed experimentally.
\end{abstract}


\end{frontmatter}

\section{Introduction}

The Wnt signalling is a conserved biochemical cell-to-cell signalling pathway controlling  embryonic development, stem cell biology and  disease, such as cancer \cite{Nusseetal.WntSignaling:2013,HopplerandMoon:2014,Nusse:2017ve}. The best understood branch of Wnt signalling, canonical or Wnt/\textbeta-catenin signalling, functions to control downstream gene expression \cite{Hoppler:2014}. During embryonic development, Wnt signalling regulates key processes, such as pluripotency, embryonic axis and germ layer induction, cell proliferation and differentiation \cite{Nusseetal.WntSignaling:2013,HopplerandMoon:2014}.    

Using \textit{Xenopus} as an experimental model organism, two sources of Wnt/\textbeta-catenin signalling were  identified in early vertebrate embryos \cite{HopplerKuhl2022}: before the onset of zygotic gene expression, also called Zygotic Gene Activation (ZGA), maternal Wnt signalling induces a dorsal axis; and then after the onset of zygotic gene expression, zygotic Wnt8 signalling promotes ventral and lateral mesoderm development; thus suggesting that we might also expect just two corresponding classes of direct Wnt target genes \cite{Hamilton:2001va,Zylkiewicz:2014,Hikasa:2013ub}. Molecular identification of direct Wnt/\textbeta-catenin target genes in recent transcriptomics and genomics studies \cite{Nakamura:2016uj,Nakamura:2017us,Afouda:2020tg}, however, now reveal considerable complexity among Wnt target genes in early development; partly due to co-regulation by other  signalling pathways, such as nodal, FGF and BMP signalling \cite{Nakamura:2016uj,Afouda:2020tg}. 

Based on these studies a useful definition of five classes of direct Wnt target genes can be proposed: two classes of direct maternal Wnt/\textbeta-catenin target genes, both of which are co-regulated by Nodal signalling, but with one class (e.g. \textit{siamois}) expressed before the other (e.g. \textit{goosecoid}) \cite{Afouda:2020tg}. A third class of apparently ubiquitous Wnt target genes is regulated by maternal and zygotic Wnt signalling (e.g. \textit{axin2}) \cite{Nakamura:2016uj,Afouda:2020tg}. As well as regulating this third class, zygotic \textit{wnt8a}/\textbeta-catenin signalling also directly regulates at least two further classes of target genes \cite{Nakamura:2016uj}; a fourth class is co-regulated by zygotic BMP signalling (e.g., \textit{msx1}) and a fifth class is co-regulated by zygotic FGF signalling (e.g., \textit{cdx2}). 

\begin{figure}[htbp]
\includegraphics[width=12cm]{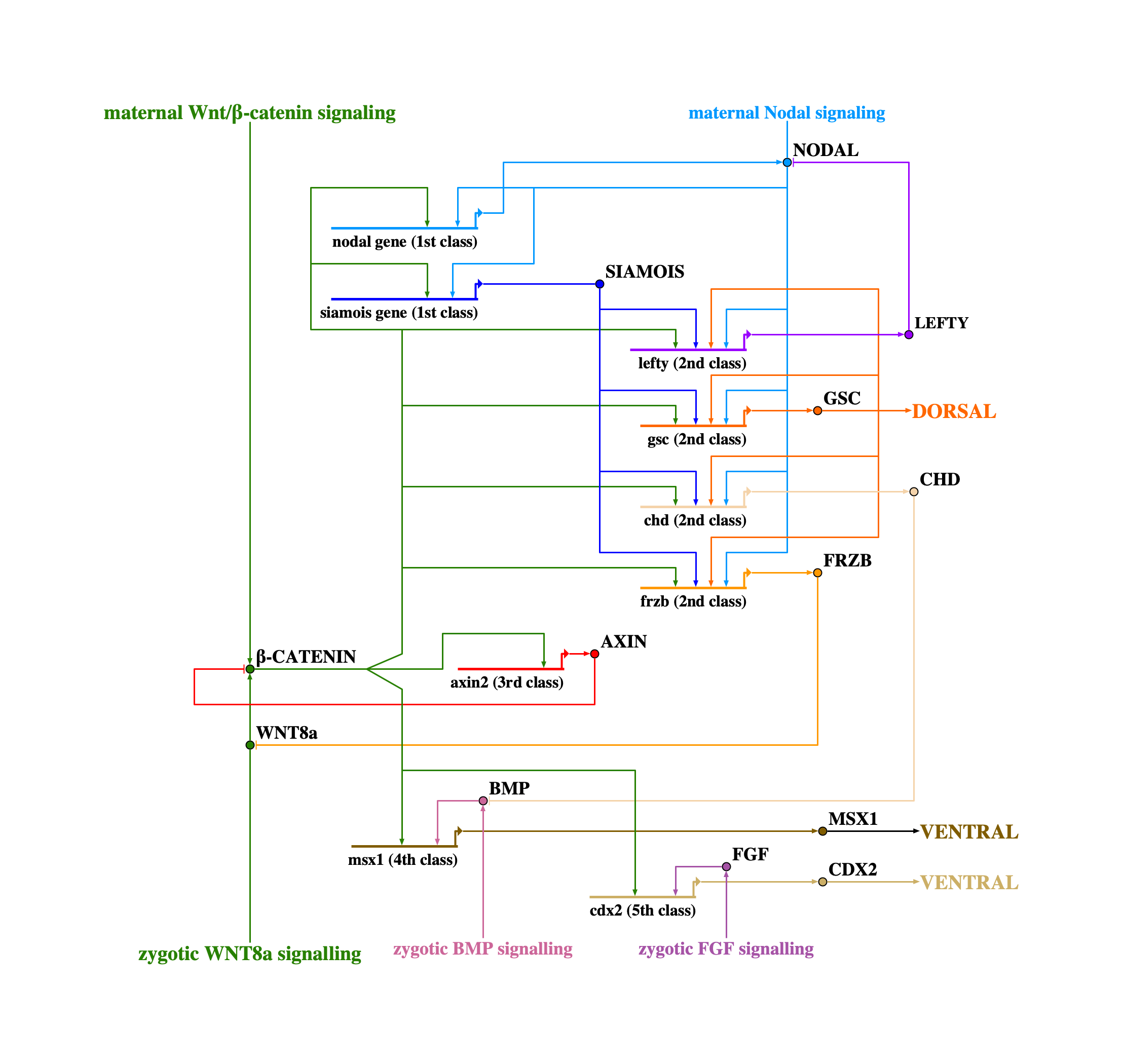}
\caption{A gene regulatory network built to represent dorsoventral axis induction in \textit{Xenopus tropicalis} embryos, with certain genes chosen from each of  five classes of direct Wnt target genes. The first class of Wnt target genes (\textit{nodal} and \textit{siamois}) are regulated by  maternal Wnt/\textbeta-catenin signalling and maternal nodal/TGF\textbeta signalling. The second class of Wnt target genes (\textit{frzb}, \textit{chordin}, \textit{goosecoid}  and \textit{lefty}) are regulated by early maternal Wnt/\textbeta-catenin signalling and products of the first class of Wnt target genes (i.e., SIAMOIS and NODAL) in a coherent, feed-forward regulatory loop manner. The third class of Wnt target genes (axin2) is regulated by both early maternal Wnt/\textbeta-catenin signalling and zygotic Wnt8a/\textbeta-catenin signalling. The fourth class of Wnt target genes (msx1) are co-regulated by zygotic BMP signalling and zygotic WNT8a/\textbeta-catenin signalling. The fifth class of Wnt target genes (cdx2) is co-regulated by zygotic FGF signalling and zygotic Wnt8a/\textbeta-catening signalling. Model constructed using BioTapestry.} 
\label{GRN1}
\end{figure}

Functional experiments have also revealed cross- and feedback regulation in this system \cite{Afouda:2020tg}: Among the maternal Wnt/\textbeta-catenin target genes, gene products of the first class (e.g. SIAMOIS, NODAL) are required for expression of genes of the second class in a coherent feed forward loop \cite{Afouda:2020tg,Mangan11980}, recently reviewed by Cho and Blitz \cite{ChoBlitz2022}. Some gene products of this second class of maternal Wnt/\textbeta-catenin target genes are known to negatively regulate zygotic \textit{wnt8}/\textbeta-catenin signalling (e.g. FRZB \cite{Wang:1997uv}) or BMP signalling (e.g. CHD, chordin \cite{Piccolo:1996vm}), which will have consequences for the expression of Wnt target genes of the forth and fifth class. Some gene products of the third class (i.e. AXIN2) regulate the Wnt\textbeta-catenin pathway in a negative feedback loop \cite{Lee:2003wv}. The complete Gene Regulatory Network model inferred from previous experimental findings \cite{Nakamura:2016uj,Afouda:2020tg} is illustrated in Figure \ref{GRN1}. However, it was not clear so far whether this diagrammatic model provides an accurate description of the experimentally observed dynamics. In an attempt to test this we have carried out so-called \textit{in silico} experiments using mathematical modelling.

Research on the WNT pathway has been minimal in terms of mathematical modelling overall. The reason for this lies in the intricacy of the pathway, the complexity of the proteins involved but more importantly, the scarcity of computable or measurable experimental data. The first representation of WNT signalling as an ODE-based model came about in 2003 by Lee et al \cite{Lee:2003wv}.  Studies introduced later have shown alternate types of modelling, such as graph representations \cite{Kestler:2008vx,Kuhl:2014} and different ways of biochemical networks \cite{Plouhinec:2011vx}. 

With the advent of high throughput sequencing it is expected that information about tens even hundreds of genes will become available \cite{Satou-Kobayashi:2021up} and classical modelling will be impractical, if not impossible. The richness of information, paradoxically, might hide fundamental types of interactions, we ``can't see the forest for the trees'' as the classical saying goes. There is therefore a need to simplify the modelling, reducing it to classes instead of individual genes.

We propose a mathematical model representing dorsoventral axis specification of \textit{Xenopus} embryos, based on ordinary differential equations (ODEs), to help elucidate the dynamics behind the Gene Regulatory Network regulating the proposed five classes of direct Wnt/\textbeta--catenin target genes that were identified in early \textit{Xenopus} embryogenesis by Nakamura, et al. and Afouda, et al. \cite{Nakamura:2016uj,Nakamura:2017us,Afouda:2020tg}. We then take this model further and show it can be simplified by introducing artificial constructs - "super-genes". This approach facilitates the understanding of the model without loosing key information. However, while the key features of the biological system are easily reproduced in our mathematical model, it also shows the need for further experiments to elucidate the expression dynamics of certain genes, even suggesting the existence of more classes than those five currently proposed.

\section{Materials and methods}

The model outline was first drawn using BioTapestry \cite{Paquette:2016vi}. In a further step, to describe the dynamics of each gene expression and function in our model we construct an ODE with a production and degradation term for each gene product. The production term includes regulation of gene expression by other components in the Gene Regulatory Network and incorporates Hill/Michaelis-Menten kinetics. Although normally used to represent enzyme kinetics, Hill/Michaelis-Menten kinetics can be adopted to model gene regulation and expression to assess the saturating effect of each reaction between the components in the gene regulatory network, and is different for activators and repressors \cite{Saka:2012ua,Blossey2008,RUSSO20095070}.

Gene expression of the  \textit{nodal} gene is regulated by maternal Nodal-like TGF\textbeta\ protein (autoregulation) and also by \textbeta-catenin while being repressed by LEFTY (Equation \ref{Nodal}). A similar gene expression equation can be written for \textit{siamois} (Equation \ref{Siamois}). Note that in order to account for differences in chromatin accessibility during the maternal and zygotic phase \cite{Nakamura:2016uj,Afouda:2020tg}, in the model the rate constants for NODAL and SIAMOIS are time dependent, modelled as logistic sigmoid functions $k_{Mat}(t)=1-\sigma (t-10)$ and $k_{Zyg}(t)=\sigma (t-10)$.

The dynamics of FRZB, GOOSECOID, CHORDIN and LEFTY are also similar , all being positively regulated by \textbeta-catenin, NODAL and GOOSECOID (Equations \ref{Frzb}, \ref{Goosecoid}, \ref{Chordin}, \ref{Lefty}).

Equation \ref{Axin2} corresponds to AXIN2 production from the \textit{axin2} gene, which is regulated only by \textbeta-catenin. Equation \ref{BCatenin} represents the production of nuclear \textbeta-catenin, whose production is regulated by WNT8a, and inhibited by AXIN2. Note that this is actually an approximation, to keep it in line with the modelling of the other variables. In reality \textbeta-catenin is constantly and ubiquitously produced and degraded, but degradation is inhibited by WNT8a, and AXIN2 reinstates degradation of \textbeta-catenin, in other words WNT8a represses the degradation, while AXIN promotes the degradation \cite{Hoppler:2014}. 

Equation \ref{BMP} corresponds to the production of BMP, which is inhibited by CHORDIN but also degraded. Equation \ref{WNT8a} is similar to Equation \ref{BMP}, except WNT8a is inhibited by FRZB. The equation \ref{FGF} is for FGF production and it simply includes constant production and degradation terms, because in our model of the gene regulatory network (Figure \ref{GRN1}), the protein is simply produced from the FGF signalling pathway. Again, note that the rate constants for BMP, WNT8a and FGF are again time dependent to account for differences in chromatin accessibility during the maternal and zygotic phase \cite{Nakamura:2016uj,Afouda:2020tg}.

Equation \ref{MSX1} represents the production of MSX1 co-regulated by \textbeta-catenin and BMP signalling while the final equation, \ref{CDX}, represents the production of CDX2, which is co-regulated by \textbeta-catenin and FGF signalling.

\begin{equation}
\label{Nodal}
NO'(t)=\frac{k_1 k_{Mat} NO(t)^h + k_2 k_{Mat} BC1(t)^h}{NO(t)^h+BC1(t)^h+L(t)^h+1}-\gamma _1 NO(t)
\end{equation}

\begin{equation}
\label{Siamois}
S'(t)= \frac{k_3 k_{Mat} NO(t)^h + k_4 k_{Mat} BC1(t)^h}{NO(t)^h+BC1(t)^h+L(t)^h+1}-\gamma _2 S(t)
\end{equation}

\begin{equation}
\label{Frzb}
F'(t)=\frac{k_5 BC1(t)^h + k_6 NO(t)^h + k_7 GSC(t)^h}{BC1(t)^h+NO(t)^h+GSC(t)^h+1}-\gamma _3 F(t)
\end{equation}

\begin{equation}
\label{Chordin}
CH'(t)=\frac{k_8 BC1(t)^h + k_9 NO(t)^h + k_{10} GSC(t)^h}{BC1(t)^h+NO(t)^h+GSC(t)^h+1}-\gamma _4 CH(t)
\end{equation}

\begin{equation}
\label{Goosecoid}
GSC'(t)=\frac{k_{11} BC1(t)^h + k_{12} NO(t)^h + k_{13}GSC(t)^h}{BC1(t)^h+NO(t)^h+GSC(t)^h+1} -\gamma_5 GSC(t) 
\end{equation}

\begin{equation}
\label{Lefty}
L'(t)=\frac{k_{14}BC1(t)^h + k_{15} NO(t)^h + k_{16} GSC(t)^h}{BC1(t)^h+NO(t)^h+GSC(t)^h+1} -\gamma_6 L(t) 
\end{equation}

\begin{equation}
\label{Axin2}
A'(t)=\frac{k_{17} BC1(t)^h}{BC1(t)^h+1}-\gamma _7 A(t)
\end{equation}

\begin{equation}
\label{BCatenin}
BC1'(t)=\frac{k_{18} WN(t)^h}{A(t)^h+WN(t)^h+1}-\gamma _8 BC1(t)
\end{equation}

\begin{equation}
\label{BMP}
BM'(t)=\frac{k_{19}k_{Zyg}}{BR(t)^h+CH(t)^h+1}-\gamma _9 BM(t)
\end{equation}

\begin{equation}
\label{WNT8a}
WN'(t)=\frac{k_{20}k_{Zyg}}{F(t)^h+1}-\gamma _{10}WN(t) 
\end{equation}

\begin{equation}
\label{FGF}
FF'(t)=k_{21}k_{Zyg} - \gamma _{11}FF(t)
\end{equation}

\begin{equation}
\label{MSX1}
M'(t)=\frac{k_{22} BC1(t)^h+k_{23} BM(t)^h}{BC1(t)^h+BM(t)^h+1}-\gamma _{12} M(t)
\end{equation}

\begin{equation}
\label{CDX}
CDX'(t)=\frac{k_{24} BC1(t)^h+k_{25} FF(t)^h}{BC1(t)^h+FF(t)^h+1}-\gamma _{13}CDX(t) 
\end{equation}

The ODEs (equations \ref{Nodal}- \ref{CDX}) were solved with the mathematics software package Mathematica 12.2 (Wolfram Research, Inc.) to produce graphs for the production of each protein as a function of time. Of particular help was the use of the \textit{Manipulate} function which allows us to visualise the effects of changes in protein rate synthesis instantly. The Mathematica codes are available in the Supplementary Material and upon request from the authors.

\section{Results}

We focus on the Gene Regulatory Network (GRN) architecture and as such we choose arbitrary values for the parameters, since it has been shown that in Wnt signalling, fold-change is important rather than absolute values \cite{Goentoro:2009vp}. However we consider that the effect of \textbeta-catenin is subtle, thus the corresponding rate constants are smaller than the others, consistent with biochemical findings \cite{Hoppler:2014}. We also assume that \textbeta-catenin is more unstable than the other proteins, resulting in the following set of parameters:
$ k_1=k_3=k_6=k_7=k_9=k_{10}=k_{12}=k_{13}=k_{15}=k_{16}=k_{18}=k_{19}=k_{20}=k_{21}=k_{23}=k_{25}=1,  k_2=k_4=k_5=k_8= k_{11}=k_{14}=k_{17}=k_{22}=k_{24}=0.1 ,\gamma_1=\gamma_2=\gamma_3=\gamma_4=\gamma_5=\gamma_6=\gamma_7=\gamma_9=\gamma_{10}=\gamma_{11}=\gamma_{12}=\gamma_{13}=0.1, \gamma_8=1, h=1.8$

 Our assumptions for parameter values are in agreement with existing literature, in both the rate constants being generally of the same order of magnitude and \textbeta-catenin having a subtle effect combined with a high degradation rate \cite{Lee:2003wv, Peshkin:2015ws, Man:2021vv}.

\begin{figure}[htbp]
\includegraphics[width=10cm]{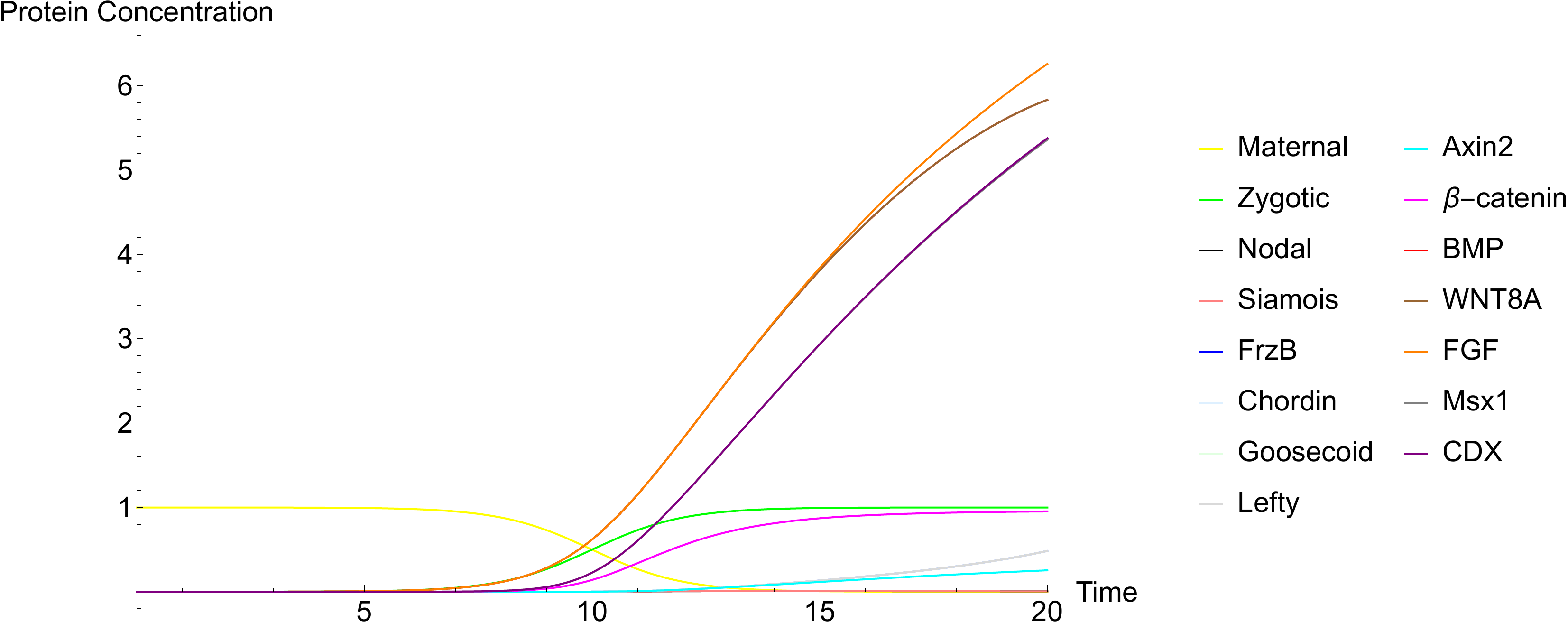}
\caption{Protein concentrations as a function of time for a mathematical model representing dorsoventral axis induction in Xenopus, for modelling ventral tissue. The units for both concentration and time are arbitrary. Note that the modelling results in simulated high expression of \textit{msx1} and \textit{cdx} and that plots for genes with similar regulation mechanisms (\textit{frzb}, \textit{chordin}, \textit{goosecoid} and \textit{lefty}) are difficult to distinguish.}  
\label{FullVentral}
\end{figure}

We consider first the ventral tissue, in which the initial concentrations for NODAL and \textbeta-CATENIN are set to 0 (Fig. \ref{FullVentral}). 

\begin{figure}[htbp]
\includegraphics[width=10cm]{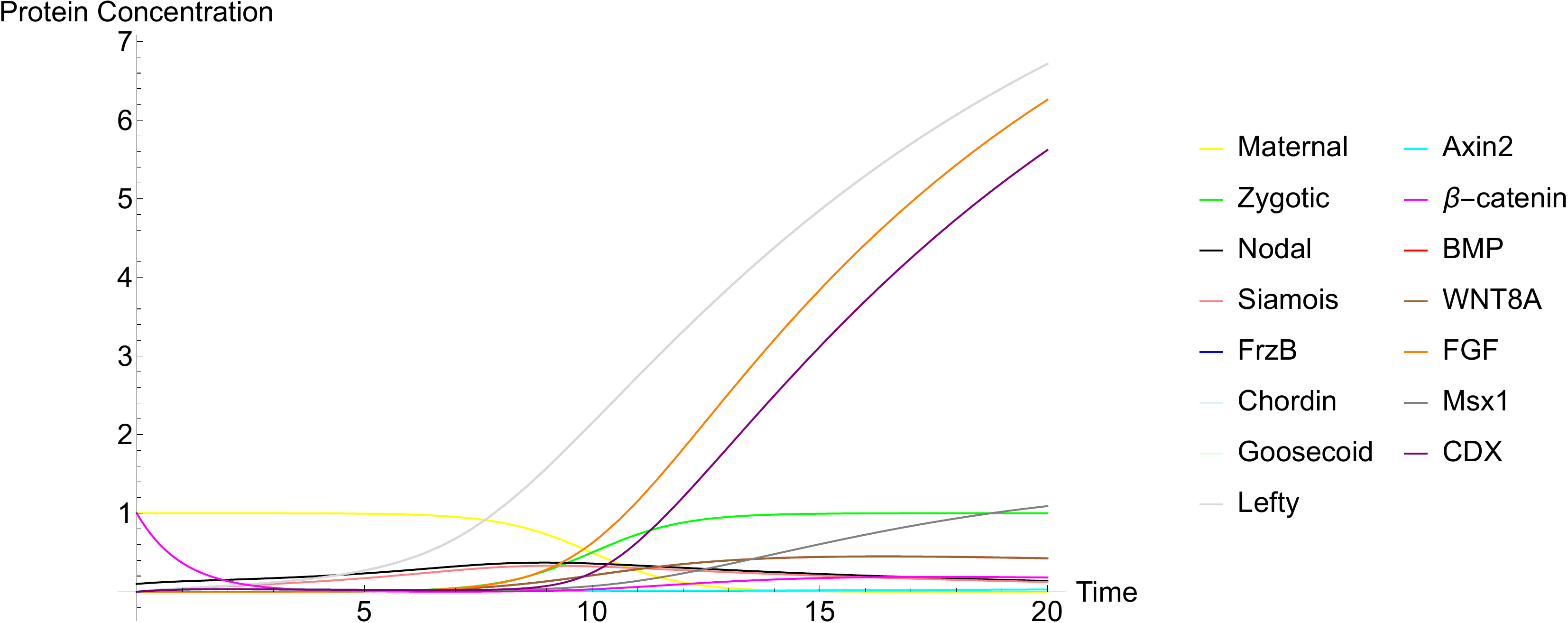}
\caption{Protein concentrations as a function of time for a mathematical model representing dorsoventral axis induction in Xenopus, for modelling dorsal tissue. Initial NODAL concentration is set to 0.1 while initial \textbeta-CATENIN concentration is set to 1. The units for both concentration and time are arbitrary. Note that the modelling results in simulated high expression of \textit{frzb}, \textit{chordin}, \textit{goosecoid} and \textit{lefty} but plots for genes with similar regulation mechanisms are difficult to distinguish.}   
\label{FullDorsal}
\end{figure}

For modelling dorsal tissue, the initial concentrations for NODAL is very small while \textbeta-CATENIN is set to high (Fig. \ref{FullDorsal}). 

The results are in agreement with the experimental findings \cite{Nakamura:2016uj,Afouda:2020tg}, however one can also easily observe in Figures \ref{FullVentral} and \ref{FullDorsal} that the dynamics of genes belonging to the same class is very similar and it is very difficult to distinguish them. The model is unnecessarily complex hence we explored whether we could simplify it.

\subsection{Simplified model with artificial conglomerate "super-genes"}

Considering only one "super-gene" per class leads to the network represented in Figure \ref{GRNsimple} and the following simplified system of ODEs:

\begin{figure}[htbp]
\includegraphics[width=12cm]{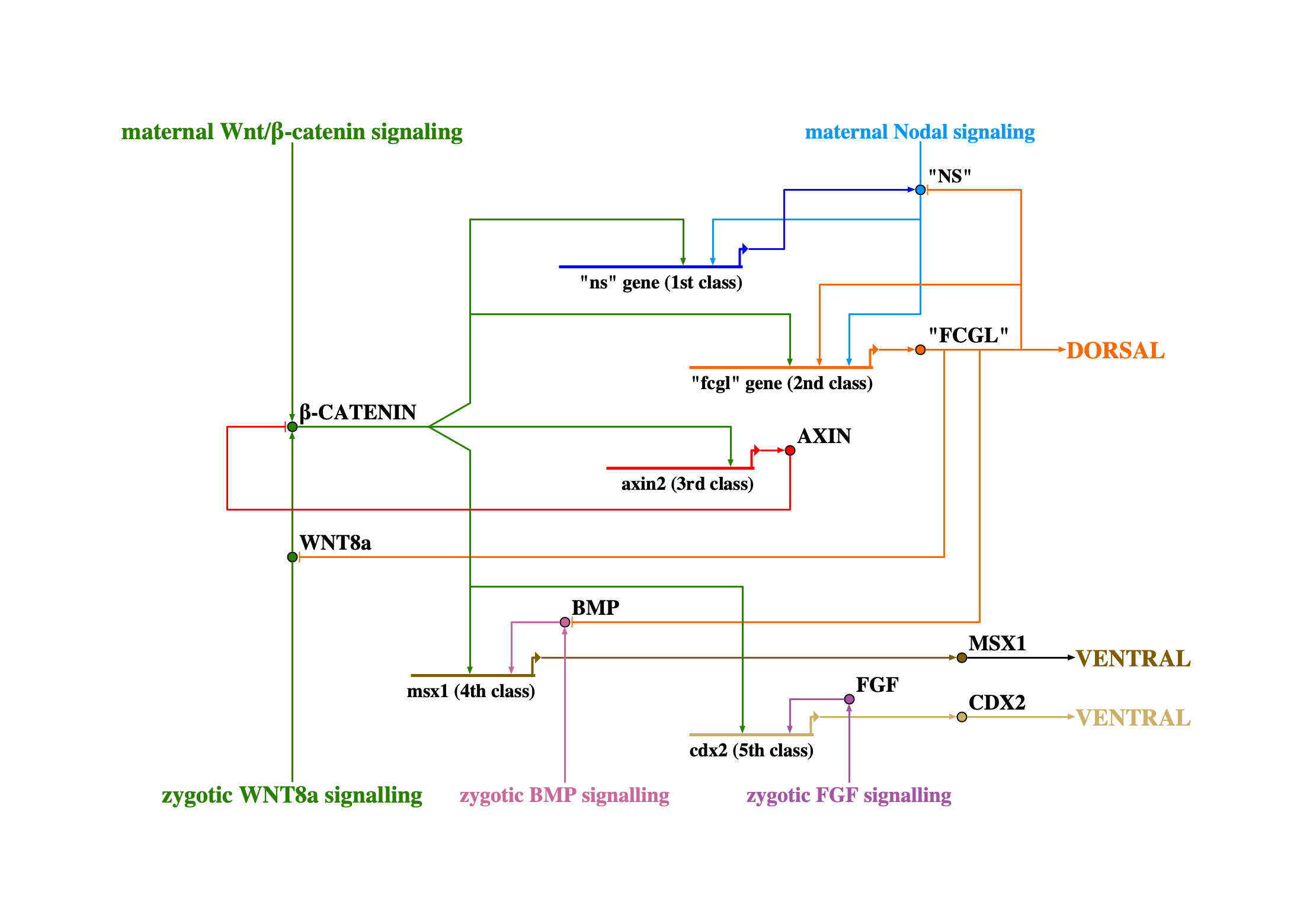}
\caption{A simplified gene regulatory network with only one "super-gene" per class. Model constructed using BioTapestry.}  
\label{GRNsimple}
\end{figure}

\begin{equation}
\label{NodalSiamois}
NS'(t)=\frac{k_1 k_{Mat} NS(t)^h + k_2 k_{Mat} BC1(t)^h}{NS(t)^h+BC1(t)^h+FCGL(t)^h+1}-\gamma _1 NS(t)
\end{equation}

\begin{equation}
\label{FCGL}
FCGL'(t)=\frac{k_5 BC1(t)^h + k_6 NS(t)^h + k_7 FCGL(t)^h}{BC1(t)^h+NS(t)^h+FCGL(t)^h+1}-\gamma _3 FCGL(t)
\end{equation}

\begin{equation}
\label{Axin2}
A'(t)=\frac{k_{17} BC1(t)^h}{BC1(t)^h+1}-\gamma _7 A(t)
\end{equation}

\begin{equation}
\label{BCatenin}
BC1'(t)=\frac{k_{18} WN(t)^h}{A(t)^h+WN(t)^h+1}-\gamma _8 BC1(t)
\end{equation}

\begin{equation}
\label{BMP}
BM'(t)=\frac{k_{19}k_{Zyg}}{BR(t)^h+FCGL(t)^h+1}-\gamma _9 BM(t)
\end{equation}

\begin{equation}
\label{WNT8a}
WN'(t)=\frac{k_{20}k_{Zyg}}{FCGL(t)^h+1}-\gamma _{10}WN(t) 
\end{equation}

\begin{equation}
\label{FGF}
FF'(t)=k_{21}k_{Zyg} - \gamma _{11}FF(t)
\end{equation}

\begin{equation}
\label{MSX1}
M'(t)=\frac{k_{22} BC1(t)^h+k_{23} BM(t)^h}{BC1(t)^h+BM(t)^h+1}-\gamma _{12} M(t)
\end{equation}

\begin{equation}
\label{CDX}
CDX'(t)=\frac{k_{24} BC1(t)^h+k_{25} FF(t)^h}{BC1(t)^h+FF(t)^h+1}-\gamma _{13}CDX(t) 
\end{equation}

\begin{figure}[htbp]
\includegraphics[width=12cm]{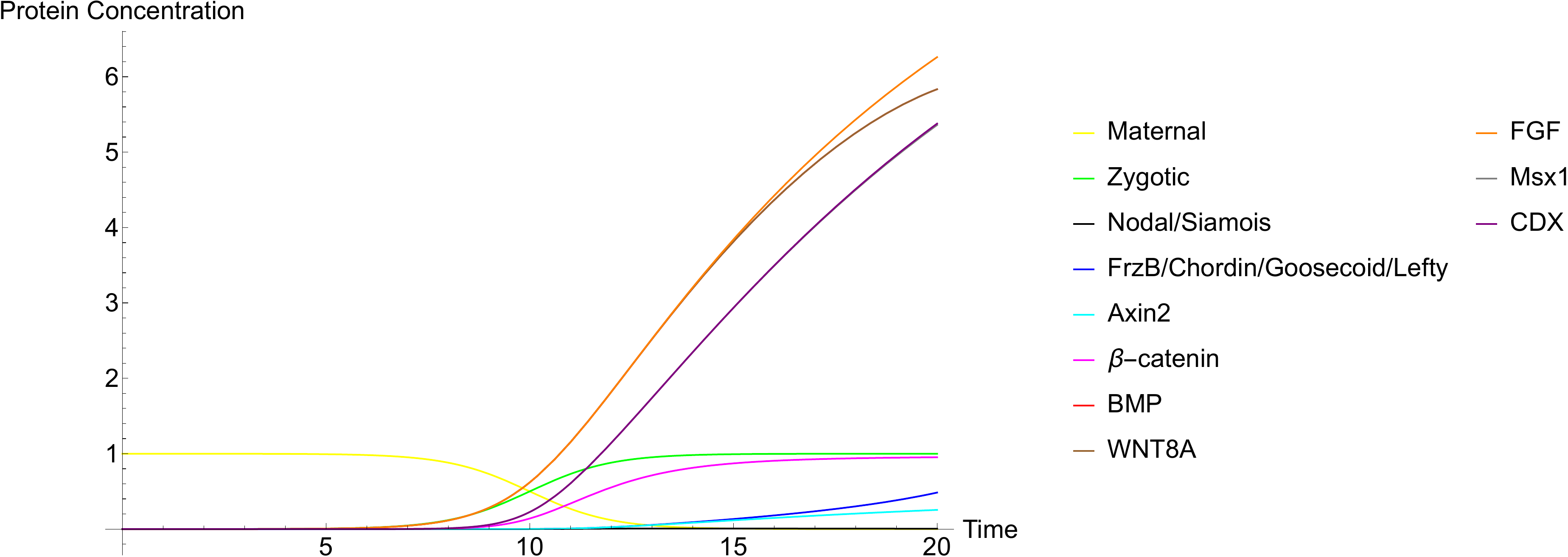}
\caption{Protein concentrations as a function of time for a simplified mathematical model with only one "super-gene' per class, in ventral tissue. The units for both concentration and time are arbitrary. Note that the modelling results in simulated high expression of \textit{msx1} and \textit{cdx}.}  
\label{SimpleVentral}
\end{figure}

\begin{figure}[htbp]
\includegraphics[width=12cm]{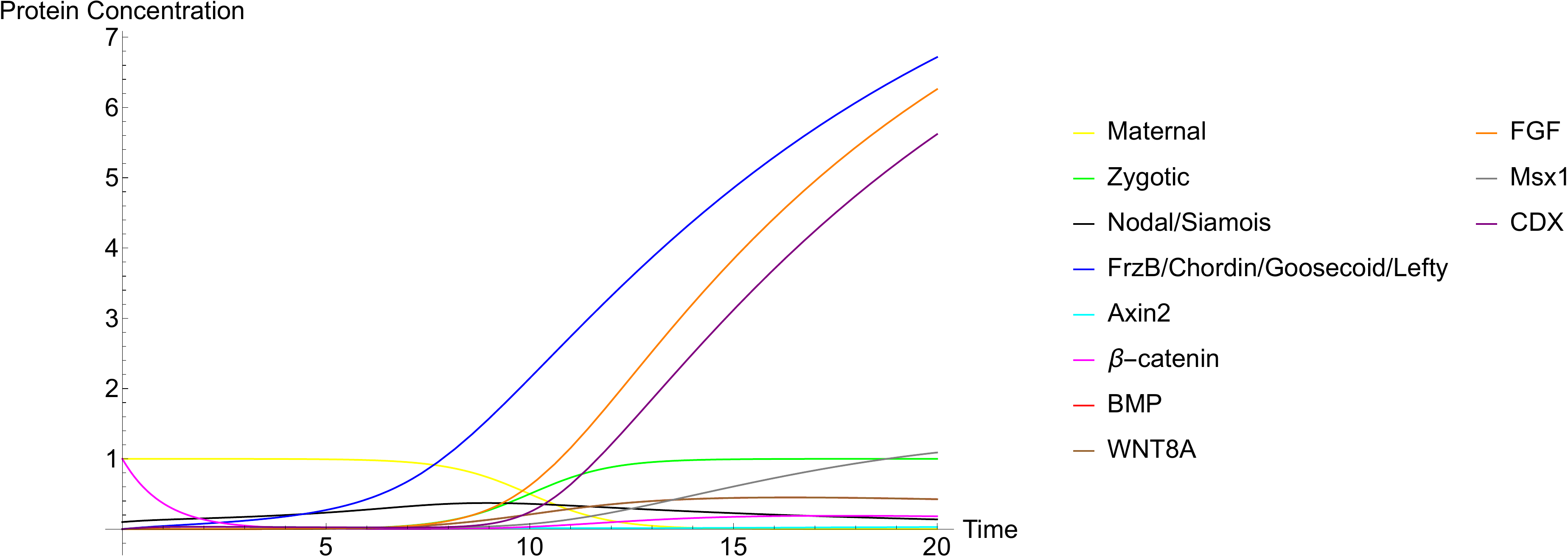}
\caption{Protein concentrations as a function of time for a simplified mathematical model with only one "super-gene' per class, in dorsal tissue. Initial NODAL concentration is set to 0.1 while initial \textbeta-CATENIN concentration is set to 1. The units for both concentration and time are arbitrary. Note that the modelling results in simulated high expression of the FrzB/Chordin/Goosecoid/Lefty "super gene".}  
\label{SimpleDorsal}
\end{figure}

While our simplified model has the advantage of facilitating the understanding of the dynamics of the system and reducing the computational load for simulations, it also has reduced utility for describing experiments where initial values or production rates of proteins produced by genes belonging to the same class are changed separately. It is thus not useful for comparing genes from the same class. Also see discussion about comparisons between class 5 genes \textit{cdx} and \textit{hoxb1}.

\subsection{Over-expression and suppression of genes}
We also studied the predictions of our model in the case of experiments involving over-expression or knock-down of genes. The use of mathematical modelling enables us to study the effect of over-activation of a gene by changing the starting value of its corresponding product, mimicking the actual experimental procedure. Similarly, the knock-down of genes can be modelled by setting the corresponding rate constants to zero. This allows us to probe the dynamics of the GRN and observe the effects that changing the amount of available protein product (or several proteins, if desired) has on the other constituents in our gene regulatory network, which allow us to test whether the model reflects the hypothesised interactions between the constituents of the gene regulatory network and, importantly, where available, the results of previous experiments. In the following \textit{in-silico} experiments we will use only the simplified model since it is sufficient to draw conclusions about the behaviour of the system.

We studied the effect of over-expression of Class I Wnt target genes (\textit{siamois} and \textit{nodal}) by setting the initial condition of the "NS" gene product to 1. These genes co-regulate the production of Spemann organiser genes products (from Class II Wnt target genes), FRZB, CHORDIN, GOOSECOID and LEFTY in a coherent feed-forward regulatory loop manner. As a result of increasing the amount of our "NS" gene products, the concentrations of the proteins of the Spemann organiser genes are also being increased resulting in a state similar to the ventral tissue, in spite of low initial \textbeta-CATENIN. 
Our modelling results confirm the correctness of the conclusions in Afouda et al. \cite{Afouda:2020tg} based on the experimental inhibition of \textit{siamois} and rescue (Fig. 3 in the above cited paper) as well as inhibition of \textit{nodal}/TGF-\textbeta{} signalling (Fig.4 K+L, ibid.).

\begin{figure}[htbp]
\includegraphics[width=12cm]{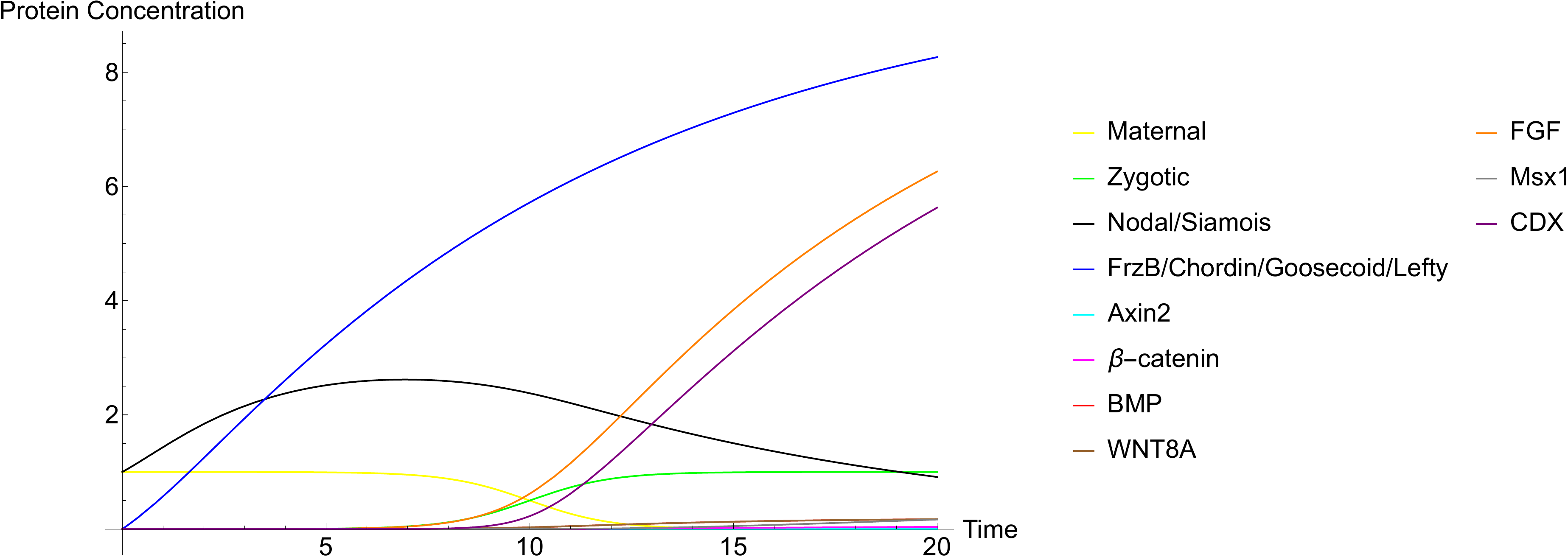}
\caption{Simulation of the over-expression of NS by setting initial NS concentration to 1 while keeping initial \textbeta-CATENIN at 0. Note that the modelling results in simulated high expression of the FrzB/Chordin/Goosecoid/Lefty "super gene".}  
\label{NodalOE}
\end{figure}

In the next step we modelled the knock-down of Class II genes. In spite of keeping the initial values of both \textbeta-CATENIN and NS high, the result is similar to ventral tissue.

\begin{figure}[htbp]
\includegraphics[width=12cm]{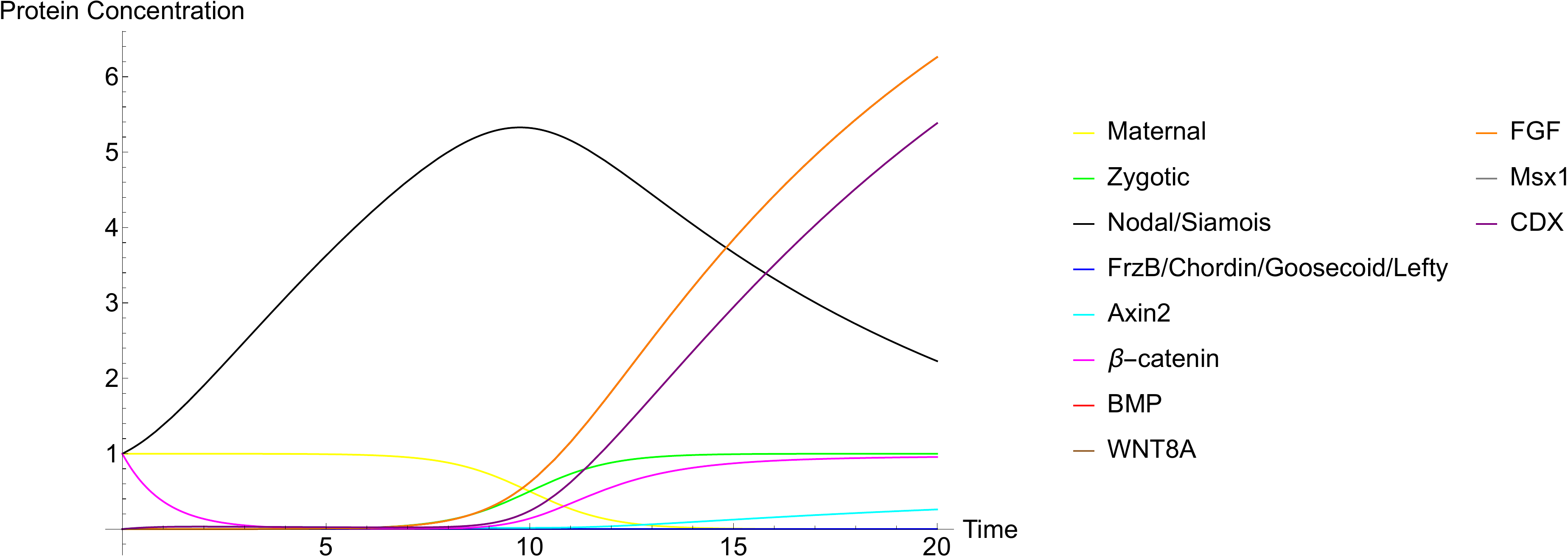}
\caption{Simulation of the knock-down of Class II genes by setting $k_5=k_6=k_7=0$ with high initial NS and \textbeta-CATENIN concentration. }  
\label{NodalOE}
\end{figure}

Contrarily, when the initial concentration of  "FCGL" is set to 1, although we start from NS and \textbeta-CATENIN at 0, the modelling essentially simulates a dorsal response with high expression of the FrzB/Chordin/Goosecoid/Lefty "super gene". This ``dorsalizing effect'' of the Class II genes has been documented previously \cite{Sasai:1994vm,De-Robertis:2001vr}.

\begin{figure}[htbp]
\includegraphics[width=12cm]{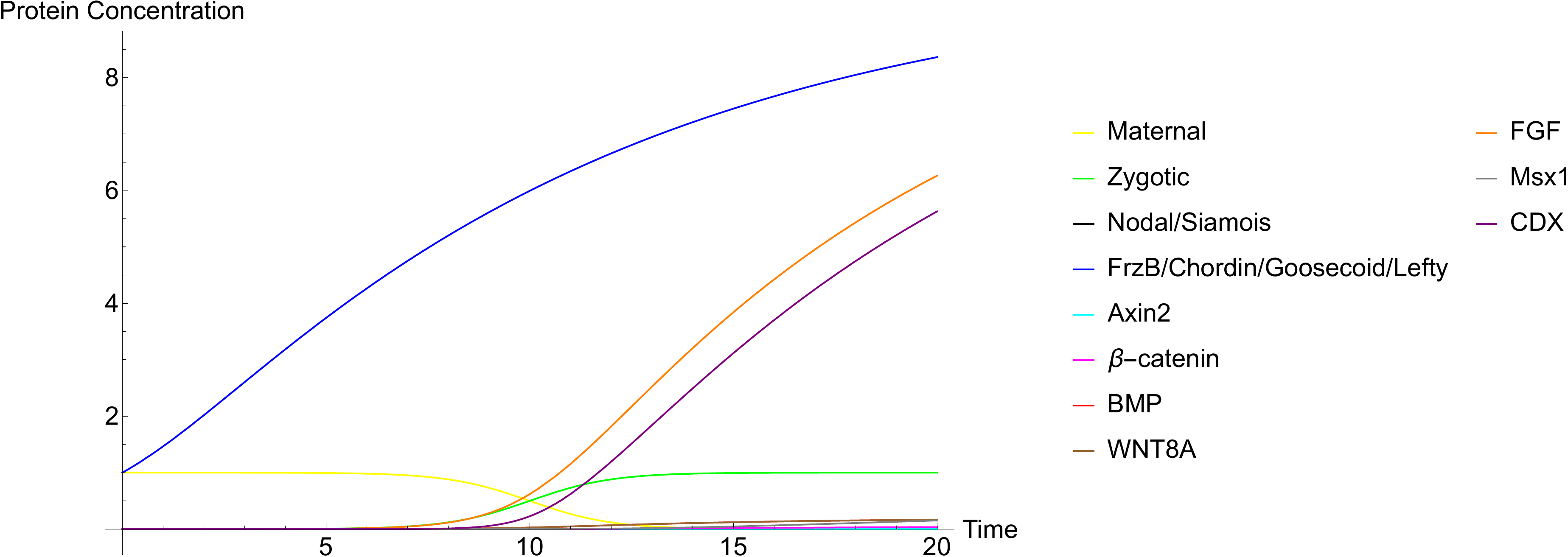}
\caption{Simulation of the experimental over-activation of Class II genes by setting initial "FCGL" concentration to 1 while keeping initial NS and \textbeta-CATENIN at 0. Note the high level of FCGL product. }  
\label{NodalOE}
\end{figure}

Experimental over-activation of BMP, by starting from initial concentration 1, also suffices for the generation of a dorsal response, although some expression of \textit{msx1} is present. The equivalent experiment had been performed by Nakamura et al. \cite{Nakamura:2016uj} and results similar to ours are presented in their Fig. 5A.

\begin{figure}[htbp]
\includegraphics[width=12cm]{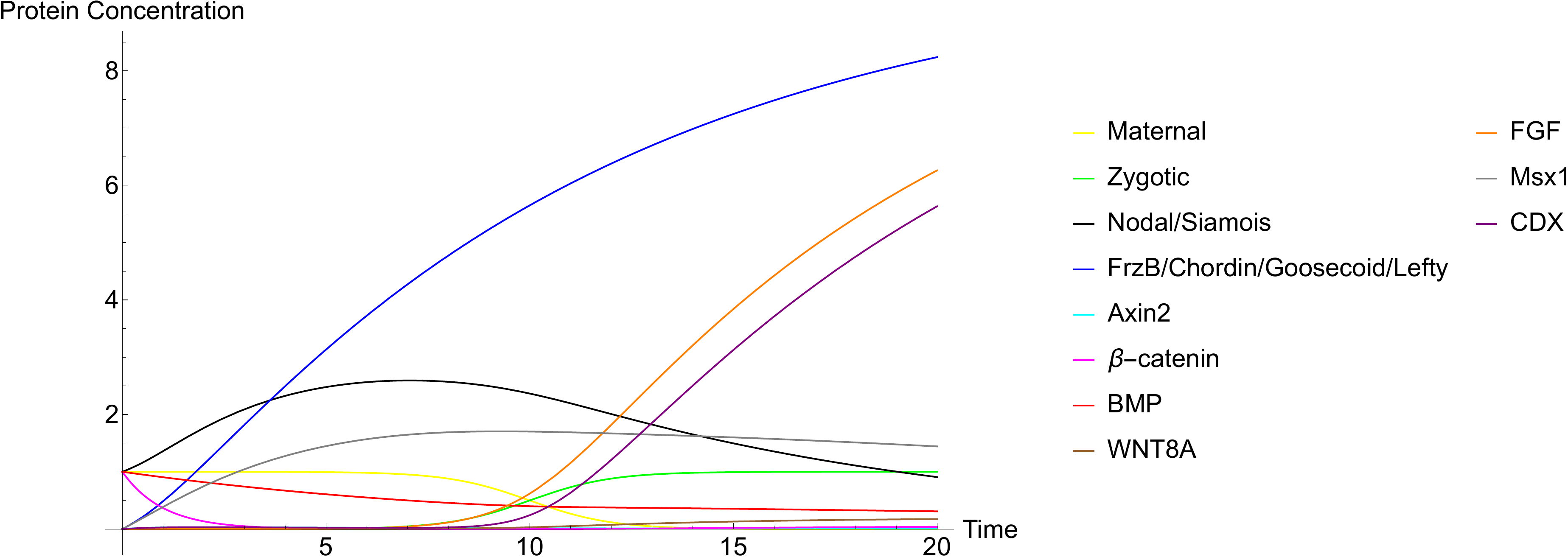}
\caption{Simulation of the experimental over-activation of BMP by setting its initial concentration to 1 with initial NODAL and \textbeta-CATENIN at 1. Note that modelling essentially simulates a dorsal response with high expression of the  FrzB/Chordin/Goosecoid/Lefty "super gene" despite some expression of \textit{msx1}. } \label{NodalOE}
\end{figure}

However, a high initial FGF level results only in a mixed response, with high expression of the FrzB/Chordin/Goosecoid/Lefty "super gene" and \textit{cdx}. High expression of \textit{cdx} when FGF is over-expressed had been observed experimentally by Keenan et al., see Fig. 2C in \cite{Keenan:2006wz}.

\begin{figure}[htbp]
\includegraphics[width=12cm]{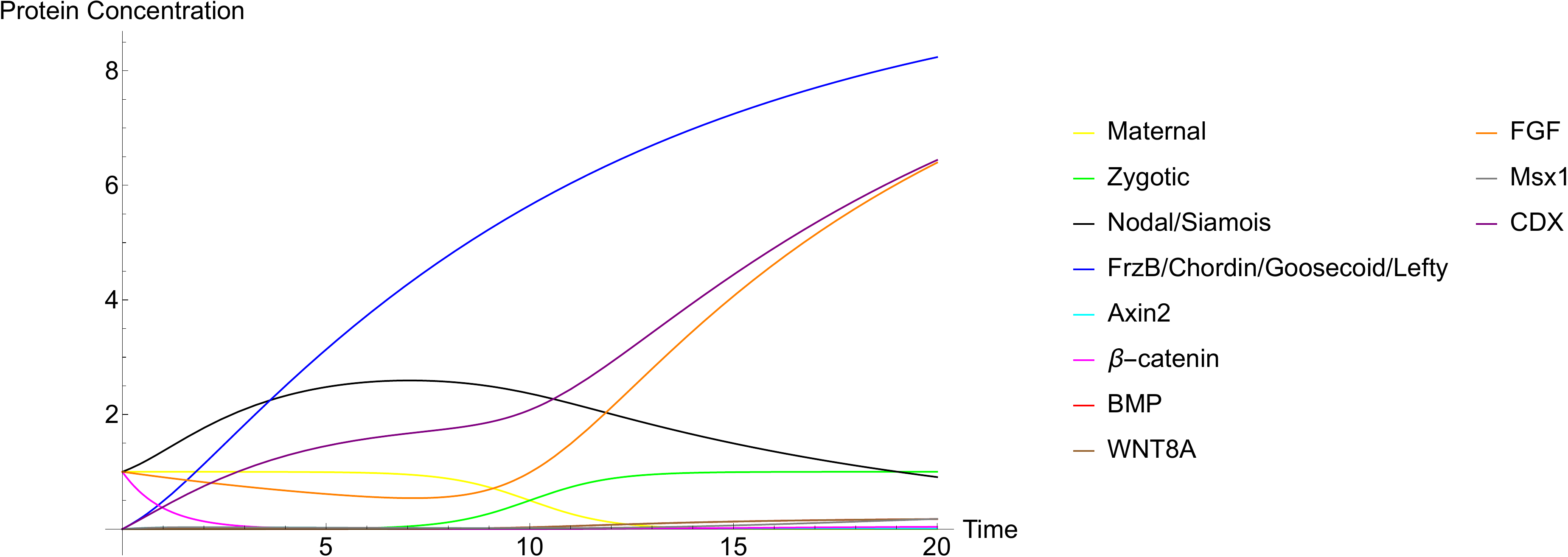}
\caption{Simulation of the experimental over-activation of FGF by setting its initial concentration to 1 with initial NODAL and \textbeta-CATENIN at 1. Note that modelling results in a mixed response with high expression of the FrzB/Chordin/Goosecoid/Lefty "super gene" and \textit{cdx}.}  
\label{NodalOE}
\end{figure}

\section{Discussion}

Molecular identification of direct target genes of Wnt/\textbeta-catenin signalling \cite{Afouda:2020tg,Nakamura:2016uj} suggests a useful definition of five classes of such Wnt target genes in early Xenopus development, which additionally regulate each other (as illustrated in Figure \ref{GRN1}). In order to test the logic and possible dynamics of this proposed Gene Regulatory Network (GRN) architecture, we created a mathematical model to describe and simulate this network. We initially picked one or two genes from each target class of Wnt genes. \textit{siamois} and \textit{nodal} were picked from class I. \textit{frzb}, \textit{chordin},\textit{goosecoid} and \textit{lefty}  were picked from class II and were referred to as the organiser genes, as they were involved in the induction of the Spemann organiser \cite{De-Robertis:2001vr}. \textit{axin2} was chosen to represent the third class of genes. \textit{msx1} represents the fourth class of genes (these were co-regulated by BMP signalling) and \textit{hoxd1} was chosen for the fifth class of target Wnt genes (this was co-regulated by FGF signalling). A gene regulatory network (Figure \ref{GRN1}) was then created to represent these genes and their respective proteins and their proposed interactions with each other and their involvement in the canonical Wnt/\textbeta-catenin signalling pathway, as suggested by the previous experiments. The context-specific mechanism, including co-regulation by nodal/TGF\textbeta/activin/Vg1, BMP and FGF signalling \cite{Nakamura:2016uj} was also a very important aspect of our model and this was incorporated into the gene regulatory network; maternal Wnt/\textbeta-catenin signalling regulated the class I, II and III genes, whereas after Zygotic Gene Activation, the zygotic Wnt8A/\textbeta-catenin signalling regulated class III, class IV and class V genes. Note that the current proposed Gene Regulatory Network is based on experiments providing snapshots of analysis of direct Wnt target genes at ZGA \cite{Afouda:2020tg} and gastrulation \cite{Nakamura:2016uj}. Analysis at other stages could suggest additional classes of direct Wnt target genes thus future research may well uncover further distinct classes of Wnt/\textbeta-catenin target genes in early Xenopus development. It should be noted that only a limited number of genes from each class of Wnt target genes was included in the gene regulatory network. Many components were missing so as not to make it unnecessarily complicated which would defeat the point of mathematically modelling this system to simplify and enhance our understanding. We anticipated that further model reduction would help us understand the new concept of five classes of Wnt target genes in a context-specific framework and it would help us create a model which is easy to understand without losing the overall understanding of the dynamics of the biology. Model reduction was achieved by excluding certain components of the gene regulatory network, excluding the separate steps of transcription and translation of genes, and the stabilisation of \textbeta-catenin (including the canonical Wnt signalling pathway destruction complex). We have therefore also followed the assumption of all gene regulation occurring at the transcriptional level and therefore ignoring here additional potential regulatory steps at the level of RNA processing or translation. 

Our model displays bistability – the system we have modelled has two states, i.e., dorsal and ventral. We first need to look into how the dorsal and ventral mesoderm is established in a Xenopus embryo. 

The unfertilised egg has an animal and vegetal pole (where nodal/TGF\textbeta-related VegT mRNA and also dorsal determinants are co-localized) and is therefore cylindrically symmetrical. This symmetry is broken post-fertilisation as when the sperm enters the egg, cortical rotation is initiated, and the dorsal determinants (namely \textbeta-catenin activators) are shifted to the prospective dorsal side of the embryo \cite{Zylkiewicz:2014}. 

The \textit{nodal} gene is activated in a higher level on the dorsal side than the ventral side of the embryo, but this still means that some nodal protein
is produced everywhere, but in varying amounts depending on the region. Where this overlap is strongest between \textbeta-catenin and \textit{nodal} is where the promoters of the organiser genes are activated, which results in accumulation of gene products of the organiser genes in the dorsal region (which in our model alludes to FrzB, \textit{chordin}, \textit{goosecoid} and \textit{lefty}). Essentially, the synergistic relationship between nodal and \textbeta-catenin leads to organiser formation \cite{Xanthos:2002vc}. This relationship is reflected in our model, such that in the dorsal response, \textbeta-catenin and nodal were set to initial conditions of 100 (see Results section). However, in the ventral response, \textbeta-catenin was set to a value of zero (see Results section). Our model assumes that the gene expression pattern (i.e., high initial \textbeta-catenin for the dorsal response, and low initial \textbeta-catenin concentration for the ventral response) supports binary cell fates – either dorsal or ventral – but in biological reality, this is not completely stable. As time progresses, permissive induction is important and further signals specify an already dorsal or ventral cell fate to complete its differentiation into, for example, the notochord. Our model therefore only considers the initial binary-like dorsoventral mesoderm gene expression, but a more elaborate model may in the future consider the later more dynamic gene expression regulating dorsoventral axis induction in \textit{Xenopus}. For example, the 5th class gene \textit{hoxb1} is more specifically expressed in the ventral tissue compared to its classmate \textit{cdx} \cite{Nakamura:2016uj} in our current model, suggesting additional regulatory mechanisms, which need further experiments to inform building of an improved model.

With our model we are able to observe the dynamics of the system based on interactions between the components in our gene regulatory network. With the system concerned here, qualitative information is more advantageous than quantitative information.

The software we chose, Mathematica, to simulate our system of ODEs, was particularly useful with the Manipulate function (see Materials and Methods), since we were able to see the effects of increasing or decreasing production rate constants on the production of components in the system instantly, which allowed us to make observations, predictions and highlight possible inconsistencies within our model. 

The model here, as with any other model representing a biological system, was based on a set of simplifying assumptions. Although this may initially appear disadvantageous, this should be viewed as a strength instead \cite{Ganusov2016}. Since no assumptions were made for the exact values of the parameters, the predictions were compared to what is currently known, allowing identification of consistencies and inconsistencies with the current knowledge. 
 A recurring and justified criticism of mathematical modelling of biological processes is the overfitting of a model with too many parameters. The fabled von Neumann's elephant can be fitted with four parameters and the fifth makes it wiggle its trunk, as proven relatively recently, more than half a century after it was stated \cite{elephant}. Note however that by imposing the rate constants of different processes to have the same value we effectively reduce the number of parameters in our model from 39 to only 5. Also our results remain valid for arbitrary variations of up to $50\%$  in the values of these parameters.
This is one of the great advantages of mathematical modelling, as it can provide insight into biological systems even where exact values of the biological parameters are difficult or currently impossible to determine, and propose experiments that can be carried out to further our understanding of the biological system.

\section{Conclusion and Perspectives}

The model created here is the first attempt at mathematical modelling of the five different classes of direct Wnt target genes recently proposed in early embryonic development \cite{Afouda:2020tg}.  It displays clear interactions of each class, providing a platform for further research in the future. 

\section{Acknowledegements}
We would like to express our gratitude to Dr Yasushi Saka for training and encouragement. Funding was received from the BHF (RG/18/8/33673 to CVG and SH) and BBSRC (BB/N021924/1; BB/M001695/1; BB/S018190/1 to SH). S.H. was a Royal Society/Leverhulme Trust Senior Research Fellow (SRF\textbackslash R1\textbackslash 191017).

\bibliography{WntModel_Hoppler}

\end{document}